\documentclass[aps, reprint, twocolumn, superscriptaddress]{revtex4-2} 
\usepackage{amssymb, amsmath, amsthm}

\usepackage{graphicx}
\usepackage{dcolumn}
\usepackage{bm}

\usepackage{dsfont}
\usepackage{color}
\usepackage{hyperref}%
\usepackage{esint}
\usepackage{float}
\usepackage{comment}

\graphicspath{{./figs/}}

\newcommand{\beq}{\begin{equation}}   
\newcommand{\eeq}{\end{equation}}
\newcommand{\beqn}{\begin{eqnarray}}   
\newcommand{\eeqn}{\end{eqnarray}}

\def\ntwo{${\mathcal N}=2\;$}

\def\none{${\mathcal N}=1\;$}
\def\ntwot{${\mathcal N}=(2,2)\;$}

\newcommand{\gsim}{\lower.7ex\hbox{$
\;\stackrel{\textstyle>}{\sim}\;$}}
\newcommand{\lsim}{\lower.7ex\hbox{$
\;\stackrel{\textstyle<}{\sim}\;$}}
\begin{document}

\title{First-order formalism for \boldmath{$\beta$} functions in bosonic sigma models from supersymmetry breaking}



\author{Oleksandr Gamayun}
\email[Correspondence to: ]{og@lims.ac.uk}
\affiliation{London Institute for Mathematical Sciences, Royal Institution,
21 Albemarle St, London W1S 4BS, UK}
\author{Andrei Losev}
\affiliation{
Shanghai Institute for Mathematics and Interdisciplinary Sciences
}%
\affiliation{
National Research University Higher School of Economics, Usacheva 6, 119048 Moscow, Russia
}
\author{Mikhail Shifman}
\affiliation{
William I. Fine Theoretical Physics Institute, University of Minnesota,
Minneapolis, MN 55455, USA
}%
 
\begin{abstract}
We consider the renormalization group flow equation for the two-dimensional sigma models with the K\"ahler target space. 
The first-order formulation allows us to treat perturbations in these models as current-current deformations. 
We demonstrate, however, that the conventional first-order formalism misses certain anomalies in the measure, and should be amended.
We reconcile beta functions obtained within the conformal perturbation theory for the current-current deformations with traditional ``geometric" results obtained in the background field methods, in this way 
resolving the peculiarities pointed out in [JHEP10(2023)097]. The result is achieved by the supersymmetric completion of the first-order sigma model.
\end{abstract}

\date{\today}

\preprint[FTPI-MINN-23-23, UMN-TH-4228/23
\maketitle


Beta functions in quantum field theories define the dependence of the coupling constant on the renormalization scale (the RG flow, see e.g. \cite{zJJ}). 
For two-dimensional sigma models, this flow has a reach geometric meaning, which is the main reason for their successful application in string theory and statistical mechanics \cite{Ketov2000}. 
In \cite{LMZ} it was shown how to cast a traditional bosonic sigma model into the
so-called first-order form, which allows one, in particular, to treat metric perturbations as conformal perturbation theory. 

In Ref. \cite{GLS} we have considered a special type of metric deformations dubbed Lie-Algebraic sigma models
\footnote{Alternative name could be bosonic Gross-Neveu models \cite{Bykov2021}}, which correspond to the current-current deformations. 
This allowed us to compare our results with the rich research history of beta functions for such deformations \cite{Kutasov1989,PhysRevLett.86.4753,Ludwig2003,Sfetsos2014,Itsios:2014lca,Georgiou:2016iom,Sagkrioti2018}.
We have established that the first-order sigma models proposed in \cite{LMZ}
when applied to $\beta$ function calculations works perfectly in the leading order. However, it leads to results incompatible with the standard geometric (background field method) calculations starting from the second order. In particular, in \cite{GLS} we considered a Lie-algebraic generalization 
of the $\mathds{C}P^1$ model on the K\"ahler space of one complex parameter $\varphi$. The metric (with the upper indices) was a finite polynomial of $\varphi\,,\bar\varphi$
 parametrized by a number of generally speaking complex parameters $n_i$. A straightforward calculation of higher loops in this formalism predicts that the higher loops must be polynomial too, which contradicts the geometric result already at two loops. 
 
We have formulated a hypothesis to explain the discrepancy as follows: 
the  loss of polynomiality in the second and higher loops is due to an infrared effect which in turn reflects the loss of symmetry in the measure not explicitly seen in the path integral.  In this work 
we will demonstrate that this is indeed the case. We consider a more general case of  K\"ahlerian target space of arbitrary dimension.
Our starting observation is as follows:
if we endow the bosonic model at hand by \ntwot supersymmetry, which is always possible, then all contributions to the $\beta$ function beyond the first loop
vanish, and simultaneously the measure is regularized. 
Next, we make superpartners' mass large and integrate them out. Remarkably, we observe a leftover --  a finite effect which can be viewed as an anomaly. 
This effect violates polynomiality. 

Our observation is somewhat similar to the situation in super-Yang-Mills (without matter). If we start from \ntwo theory, its perturbative $\beta$ function
contains only one loop which does not violate holomorphy in the complexified gauge coupling $1/g^2$. Now, if we add the mass term to the scalar superfield (including the ``second" gluino) we reduce \ntwo down to \none
breaking holomorphy starting from the second loop \cite{NSVZ,SV}
as a result of an anomaly in the measure \cite{AHM}.

Technically,  the failure of the first-order formalisms in higher loops in \cite{GLS} and the successful resolution which will be reported below is due to the following circumstance. In \cite{GLS}
the original $\beta\gamma$ system was defined classically, i.e. with the flat metric, while our perturbation used a curved metric. 
Now, through additional supersymmetry and heavy fermion masses we obtain the bosonic $\beta\gamma$ system at the quantum level, 
taking into account a non-flat metric in the measure.

More specifically, by the bosonic action of the unperturbed $\beta\gamma$ system, we understand the following sigma-model
\begin{equation}\label{S0}
	S_0 = \int_\Sigma \frac{d^2z}{\pi}  \left(
	p_a \bar{\partial}\varphi^a + \bar{p}_{\bar{a}}\partial \bar\varphi^{\bar{a}}
	\right).
\end{equation}
Here the scalar fields 
 $\varphi^a$ and $\bar\varphi^{\bar{a}}$ represent coordinates of the D-dimensional target space\footnote{We presume that $D$ is even and that indices $a$ and $\bar{a}$ runs from $1$ to $D/2$}.  The fields $p_a$, $p_{\bar{a}}$ are $(1,0)$  and $(0,1)$ forms on $\Sigma$ correspondingly.
For our purposes, it will be enough to consider $\Sigma = \mathds{C}P^1$.
This theory is classically invariant under the diffeomorphism transformations $\varphi^a \to \varphi^a - \epsilon V_A^a(\varphi)$, generated by the current $J_A = p_aV_A^a$. 
On the quantum level this symmetry becomes \textit{anomalous} which is reflected in the following operator product expansion (OPE) of the currents
\begin{equation}\label{jj1}
	J_A(z)J_B(0) = -\frac{\eta_{AB}}{z^2}+ \frac{f^C_{AB}J_C(0)- \Omega_{AB}}{z} + {\rm reg}. 
\end{equation}
Here we assume that the vector fields form an algebra with the structure constants $f_{AB}^C$.
The structures $\eta_{AB}$ and $\Omega_{AB}$ in a chosen coordinate frame are
\begin{equation}\label{etaAB}
	\eta_{AB} = \partial_k V^c_A \partial_c V_B^k,\qquad 
	\Omega_{AB} = \partial_c V_B^k\partial_m\partial_k V^c_A  \partial \varphi^m.
\end{equation}	
In the general case, both $\eta$ and $\Omega$ are functions on the target space (depending on both $\varphi$ and  $\bar{\varphi}$). The operator product expansion \eqref{jj1} does not represent a chiral \textit{current} algebra. Moreover, it is not even a
vertex operator algebroid studied in \cite{Malikov1999,malikov1999chiral,nekrasov2005lectures}, because of $\bar{\varphi}$ dependence.
This way, we have a \textit{new structure} not encountered before. 
However, in the general case $\eta_{AB}$ depends on the choice of the coordinate frame and does not transform as a scalar under the diffeomorphisms of the target (similar to  \cite{Malikov1999,malikov1999chiral,nekrasov2005lectures}). Therefore, it is even more surprising that under the current-current deformation of the theory $S_0$, specified by the action $S_G$
\begin{equation}\label{def1}
    S_G = \int \frac{d^2z}{\pi}G^{A\bar{A}} J_{A}(\varphi)J_{\bar{A}}(\bar{\varphi})
\end{equation}
the corresponding beta function (understood as a flow of the "couplings" $G^{A\bar{A}}$ ) in the first two loops reads as  
\begin{equation}\label{beta22}
	[\beta^{\rm algebra}_2]^{A\bar{A}} = \frac{1}{2} G^{B\bar{B}} G^{C\bar{C}} f^A_{BC}f^{\bar{A}}_{\bar{B}\bar{C}},
\end{equation}
\begin{equation}\label{beta33}
[\beta^{\rm algebra}_3]^{A\bar{A}} = \frac{\alpha'}{2}G^{C\bar{C}}G^{B\bar{B}}G^{F\bar{A}} \left( f_{CF}^Df_{BD}^A\eta_{\bar{C}\bar{B}}  + {\rm c.c.}	\right).
\end{equation}

The subscripts of the beta functions correspond to the power of the perturbation operators (in conformal perturbation methods \cite{GLS}) and have to be identified with a number of loops $+1$. For instance, $\beta_2^{\rm algebra}$ corresponds to the first loop, $\beta_3^{\rm algebra}$ to the second, and so on.
The superscript indicates that these expressions can be obtained solely using the current \textit{algebra} \eqref{jj1}. 
Additionally we have restored parameter $\alpha'$. 
In Ref. \cite{GLS} expressions \eqref{beta22} and \eqref{beta33} were obtained in a more general version of the deformation \eqref{def1}
\begin{equation}\label{def2}
    S_g = \int \frac{d^2z}{\pi} g^{a\bar{a}}(\varphi,\bar{\varphi}) p_a\bar{p}_{\bar{a}}.
\end{equation}
 In this case the integration over $p_a$ and $p_{\bar{a}}$ transforms the first-order sigma model into the traditional second-order geometric representation \cite{LMZ}.  
 This allows us to compare two beta functions. 
Specifically, in the first two loops  for a generic K\"ahler metric $g^{a\bar{a}}$, the corresponding beta function is well defined via the geometric objects of the target space \cite{Friedan1980,Friedan1985}
\begin{equation}\label{geom}
\beta^{\rm geometry}_2 = R^{a\bar{a}}, \qquad \beta^{\rm geometry}_3 = \frac{\alpha'}{2}R^a\,_{m\bar{p}b} R^{\bar{a}m\bar{p}b}.
\end{equation}

One can show \footnote{See Sec. S1 of the supplementary material}  that 
expressions for the one-loop beta functions do coincide, 
\begin{equation}
    \beta_2^{\rm algebra} = \beta_2^{\rm geometry},
\end{equation}
after the proper identification of $g^{a{\bar{a}}}$ and $G^{A\bar{A}}$ via equating \eqref{def1} and \eqref{def2}. 
This can be attributed to the fact that the structure constants $f_{AB}^C$ do not require any extra structures to be defined. 
In the two-loop case (i.e. for $\beta_3$) this is no longer true and extra care is needed to address the special nature of the structure $\eta_{AB}$ mentioned above. 
Alternatively, this can be considered as subtleties of integration over momenta $p_a$ and $\bar{p}_{\bar{a}}$ (see \cite{Schwarz1993,Buscher1987,Buscher1988,Hoare2019}).  

To avoid these subtleties and reconcile algebraic and geometric answers we introduce a supersymmetric generalization of the original $\beta\gamma$ system. 
Namely, we introduce fermions $\pi_a$ ($\bar{\pi}_{\bar{a}}$) and $\psi^a$ ($\bar{\psi}^{\bar{a}}$) and modify $S_0$ as
\begin{equation}
S_0 \to S_0 - \int \frac{d^2z}{\pi} \left(\pi_a \bar{\partial} \psi^a+ \bar{\pi}_{\bar{a}} \partial \bar{\psi}^{\bar{a}}\right).
\end{equation}
The  currents $J_A$ are promoted to the supersymmetric ones 
 \begin{equation}
 	J_A \to \mathcal{J}_A = p_a V_A^a(\varphi) - \pi_a \partial_b V_A^a \psi^b.
 \end{equation}
Their current algebra is no longer anomalous 
\begin{equation}
\mathcal{J}_A(z)\mathcal{J}_B(0) =   \frac{f_{AB}^C \mathcal{J}_C(0)}{z} + {\rm reg}. 
\end{equation} 
To be able to mode out the fermions we additionally introduce fermionic mass terms  
\begin{equation}\label{deltaS}
	\delta S_m =  m \int    \frac{d^2z}{2\pi} G^{a\bar{a}}\pi_a \bar{\pi}_{\bar{a}} +m \int  \frac{d^2z}{2\pi}  G_{\bar{a}a} \bar{\psi}^{\bar{a}}\psi^a.
\end{equation}
So far we do not require any symmetric properties of the matrices $G_{a\bar{a}}$ and $G^{\bar{a}a}$ and their relation to the deformations \eqref{def2}, although we assume for simplicity that they are inverse of each other 
\begin{equation}\label{inverse}
	G_{a\bar{a}}G^{\bar{a}b} = \delta_a^b,\qquad 
	G^{\bar{b}a}G_{a\bar{b}} = \delta_{\bar{a}}^{\bar{b}}.
\end{equation}
The mass-deformed action $S_m = S_0 + \delta S_m$ is invariant with respect to the diffeomorphisms generated by the vector field $V_A^a(\varphi)$ provided 
that the deformations transform \textit{covariantly},
\begin{align}\label{transf}
	\delta_A G_{\bar{a}a}  &=V_A^b \partial_b G_{\bar{a}a} + G_{\bar{a}b} \partial_a V_A^b,\\
	\delta_A G^{a\bar{a}}  & =V_A^b \partial_b G^{a\bar{a}} - G^{b\bar{a}} \partial_b V_A^a.
\end{align}
The quantity $\eta_{AB}$ previously defined via the OPE \eqref{jj1} can be alternatively defined via the two-point function of the currents. Indeed, in the purely bosonic theory, we have
\begin{equation}\label{eta0}
	\langle J_A(z) J_B(0)\rangle_{S_0} = -\frac{\partial_a V_A^b\partial_b V_B^a}{z^2}  = - \frac{\eta_{AB}}{z^2}.
\end{equation} 
One might expect that in the supersymmetric version with the mass-deformation switched on, all fermionic degrees of freedom decouple at the distances $m|z|\gg1$, and we immediately recover the first expression in Eq. (\ref{etaAB}),
\begin{equation}
	\langle \mathcal{J}_A(z) \mathcal{J}_B(0)\rangle_{S_m} \overset{m|z|\gg 1}{=} -\frac{\partial_a V_A^b\partial_b V_B^a}{z^2}.
\end{equation} 
However, this would be a hasty conclusion. It turns out to be true only for the constant matrices $G_{a\bar{a}}$. 

We claim that the derivatives of $G_{a\bar{a}}$ modify this expression already in the leading order in $m|z|$ by introducing  the covariant derivatives associated with $G_{a\bar{a}}$ (instead of the partial derivatives as in \eqref{eta0}),
\begin{equation}\label{OPE2}
	\langle  \mathcal{J}_A(z)  \mathcal{J}_B(0)\rangle_{S_m} \overset{m|z|\gg 1}{=} -\frac{\nabla_a V_A^b\nabla_b V_B^a}{z^2} .
\end{equation} 
Here 
\begin{equation}\label{conn}
	\nabla_a V_A^b = \partial_a V_A^b + G^{b\bar{a}} \partial_c G_{\bar{a}a} V^c_A. 
	\end{equation}
	Notice that this object transforms as a tensor upon the diffeomorphisms shown in \eqref{transf}. So, $\eta_{AB}$ is now a proper scalar
	\begin{multline}
	\eta_{AB} =\nabla_a V_A^b\nabla_b V_B^a= 
 \partial_a V_A^b\partial_b V_B^a +  G^{b\bar{a}} \partial_c G_{\bar{a}a} V^c_A\partial_b V^a_B \\+  G^{b\bar{a}} \partial_c G_{\bar{a}a} V^c_B\partial_b V^a_A  -
	V^b_B V^a_A\partial_aG_{\bar{c}c} \partial_b G^{c\bar{c}} \,. 
	\end{multline}
The outline of the derivation of this statement is presented in the supplementary material (Sec. S3), where 
we rigorously derive terms linear in the derivatives, while quadratic terms are recovered in the special perturbative regime. 
 
This connection reminds the Hermitian connection 
\begin{equation}\label{connH}
	\nabla^H_a V_A^b = \partial_a V_A^b + G^{b\bar{a}} \partial_a G_{\bar{a}c} V^c_A. 
\end{equation}
However, the Hermitian connection  is compatible with the metric 
\begin{equation}
	\nabla^H_aG^{b\bar{k}} = 0,
\end{equation}
while for our connection we have 
\begin{equation}
	\nabla_a G^{b\bar{k}} = \partial_a G^{b\bar{k}} + G^{b\bar{a}} \partial_c G_{\bar{a}a}  G^{c\bar{k}}.
\end{equation}
This is non-zero unless the metric is K\"ahlerian. For the K\"ahlerian metrics the both connections reduce to the Levi-Civita connection. 
Moreover, one can demonstrate\,\footnote{See Section S2 of the Supplementary material} that for the case when the fermions mass metric coincides with the deformed metric $G^{a\bar{a}} = g^{a\bar{a}}$  the algebraic beta function 
\eqref{beta33}
coincides with the geometric one \eqref{geom}
\begin{equation}
    \beta_3^{\rm algebra} = \beta_3^{\rm geometry}.
\end{equation}

In particular, in the example considered in \cite{GLS}, Eq. (2.52) -- one-dimensnional K\"ahler space (a Lie-algebraic generalization of $\mathds{C}P^1$) -- we find
\begin{equation}
    \beta_{G_{1\bar 1}} = G_{1\bar 1}\left[\frac{1}{4\pi}{\mathcal R}+ \left(\frac{1}{4\pi}{\mathcal R}\right)^2 +...
    \right]
    \label{24}
\end{equation}
where $\mathcal R$ is the scalar curvature and $\alpha'=1/(2\pi)$. Our conjecture amounts to summing the geometric progression in Eq. (\ref{24}.
The third and higher-order loops are scheme dependent, however. We plan to address this issue in the subsequent publication.

{\em Conclusion}. We conclude with the statement that the first-order formalism in 2D sigma models must be amended to take into account an anomaly in the measure, as was conjectured in Ref. \cite{GLS}. In this work we present the proof of this fact in a particular regularization, namely,
supersymmetry-based 
regularization. We calculated the second loop of the beta function using thus regularized 
first-order formalism. We demonstrated that tending the fermion masses to infinity leaves a finite trace in the bosonic model. The above residual non-vanishing contribution amends the first-order formalism result and makes it identical to the ``geometric" calculation. A new understanding gained in this study is uncovering the anomalous nature of the second and higher loops. This indicates that the exact all-order beta function most probably can be recovered on K\"ahlerian target spaces.

{\em Acknowledgments}. 
We are grateful to Nat Levine for illuminating discussions and to Vyacheslav Lysov for careful reading of the manuscript and numerous remarks. 

The work of MS is supported in part by DOE grant DE-SC0011842. Andrey Losev was supported by the Basic Research Program of the National Research University Higher School of Economics.

\bibliography{litra}
\end{document}


\title{Supplemental Material\\First-order formalism for $\beta$ functions in bosonic sigma models from supersymmetry breaking}





\maketitle

\appendix

\setcounter{equation}{0}
\setcounter{figure}{0}
\setcounter{table}{0}
\makeatletter

\renewcommand{\theequation}{\thesection.\arabic{equation}}
\renewcommand{\thefigure}{S\arabic{figure}}

\renewcommand{\appendixname}{}

\renewcommand{\thesection}{S\arabic{section}}
\titleformat{\section}[hang]{\large\bfseries}{\thesection}{0.5em}{}
\titleformat{\subsection}[hang]{\bfseries}{\thesection.\thesubsection}{0.5em}{}
\def\p@subsection     {\thesection.}

\tableofcontents

\section{One-loop comparison}

Let us demonstrate the equivalence of the geometric and algebraic beta function for the K\"ahler metric in one loop level. 
First we transform the algebraic beta function 
\begin{equation}\label{beta22}
    [\beta^{\rm algebra}_2]^{C\bar{C}} = \frac{1}{2} G^{A\bar{A}}G^{B\bar{B}} f^C_{AB}f^{\bar{C}}_{\bar{A}\bar{B}}
\end{equation}
in the form of the metric $g^{a\bar{a}} = G^{A\bar{A}}V_A^a \bar{V}_{\bar{A}}^{\bar{a}}$. Contracting the beta function in Eq. \eqref{beta22} with the $V_C^a \bar{V}_{\bar{C}}^{\bar{a}}$ we obtain 
\begin{equation}
    [\beta^{\rm algebra}_2]^{a\bar{a}} =     [\beta^{\rm algebra}_2]^{C\bar{C}} V_C^a \bar{V}_{\bar{C}}^{\bar{a}} = \frac{1}{2} G^{A\bar{A}}G^{B\bar{B}} V_{[A}^k \partial_k V^a_{B]}\bar{V}_{[\bar{A}}^{\bar{k}} \partial_{\bar{k}} \bar{V}^{\bar{a}}_{\bar{B}]} = G^{k\bar{k}}\partial_k\partial_{\bar{k}} G^{a\bar{a}} - \partial_{\bar{k}} G^{k\bar{a}} \partial_k G^{a\bar{k}}.
\end{equation}
The geometric beta function is given by the Ricci tensor
\begin{equation}
    [\beta^{\rm geometry}_2]^{a\bar{a}} =  R^{a\bar{a}} 
\end{equation}
For the K\"ahler metric $G$ we can use the following definition of the Ricci tensor 
\begin{equation}
    -R_{i\bar{j}} = \partial_i\partial_{\bar{j}} \log (G) =  \partial_i (G^{k\bar{l}} \partial_{\bar{j}} G_{k\bar{l}})  =  
    \partial_i G^{k\bar{l}} \partial_{\bar{j}} G_{k\bar{l}} +G^{k\bar{l}} \partial_i \partial_{\bar{j}}G_{k\bar{l}}  = 
    \partial_i G^{k\bar{l}} \partial_{\bar{l}} G_{k\bar{j}} +g^{k\bar{l}} \partial_i \partial_{\bar{j}}G_{k\bar{l}} 
\end{equation}
where in the last line we have used the K\"ahler property. 
One can easily prove the following lemma 
\begin{equation}
\partial_k\partial_{\bar{l}} G^{a\bar{a}} + G^{i\bar{a}} G^{a\bar{j}} \partial_k\partial_{\bar{l}} G_{i\bar{j}} 
+ G^{i\bar{a}} \partial_{\bar{l}} G_{i\bar{j}} \partial_k G^{a\bar{j}} + 
G^{i\bar{a}} \partial_k G_{i\bar{j}} \partial_{\bar{l}} G^{a\bar{j}} = 0,
\end{equation}
which allows us to show that 
\begin{equation}
    R^{a\bar{a}} = G^{k\bar{l}} \partial_k\partial_{\bar{l}} G^{a\bar{a}} - \partial_{\bar{j}} G^{k\bar{a}} \partial_k G^{a\bar{j}}.
\end{equation}
This way, we conclude that
\begin{equation}
    \boxed{\beta_2^{\rm algebra} = \beta_2^{\rm geometry}.}
\end{equation}

\section{Two-loop comparison}

In this section, we compare an algebraic and geometric beta function at two-loop level. 
The algebraic beta function reads 
\begin{equation}
    [\beta^{\rm algebra}_3]^{E\bar{C}} = \frac{1}{2}G^{A\bar{A}}G^{B\bar{B}}G^{C\bar{C}} f_{AC}^Df_{BD}^E\eta_{\bar{A}\bar{B}}  + {\rm c.c.}	
\end{equation}

Let us present this expression in terms of the metric of the target space, which for simplicity, is assumed to be (i) K\"ahler; (ii) equivalent to the metric of the mass deformation. 

We start by rewriting identically 
\begin{multline}
G^{A\bar{A}}G^{B\bar{B}}G^{C\bar{C}}  f_{\bar{A}\bar{C}}^{\bar{D}} f_{\bar{B}\bar{D}}^{\bar{E}} \eta_{AB}V_{C}^a\bar{V}_{\bar{E}}^{\bar{a}} = 
G^{A\bar{A}}G^{B\bar{B}}G^{C\bar{C}} \left(
V_{\bar{B}}^{\bar{k}} V_{\bar{A}}^{\bar{s}}\partial_{\bar{k}}\partial_{\bar{s}}  V_{\bar{C}}^{\bar{a}} +
V_{\bar{B}}^{\bar{k}} \partial_{\bar{k}}V_{\bar{A}}^{\bar{s}}\partial_{\bar{s}}  V_{\bar{C}}^{\bar{a}}-
V_{\bar{B}}^{\bar{k}} \partial_{\bar{k}} V_{\bar{C}}^{\bar{s}}\partial_{\bar{s}}  V_{\bar{A}}^{\bar{a}}   - 
V_{\bar{B}}^{\bar{k}} V_{\bar{C}}^{\bar{s}}\partial_{\bar{k}}  \partial_{\bar{s}}  V_{\bar{A}}^{\bar{a}} \right. \\
\left.
+V_{\bar{C}}^{\bar{k}} \partial_{\bar{k}} V_{\bar{A}}^{\bar{s}}\partial_{\bar{s}}  V_{\bar{B}}^{\bar{a}} 
-V_{\bar{A}}^{\bar{k}} \partial_{\bar{k}} V_{\bar{C}}^{\bar{s}}\partial_{\bar{s}}  V_{\bar{B}}^{\bar{a}} 
\right)
\left(\partial_k V_A^b\partial_b V_B^k +  G^{b\bar{c}} \partial_c G_{\bar{c}k} V^c_A\partial_b V^k_B +  G^{b\bar{c}} \partial_c G_{\bar{c}k} V^c_B\partial_b V^k_A  -
	V^b_B V^k_A\partial_kG_{\bar{c}c} \partial_b G^{c\bar{c}}  \right)V_{C}^a
\end{multline}
Here we have used the definition of the commutator of the vector fields and the modified definition of $\eta_{AB}$ (See (21) in the main text).  
Using $A\leftrightarrow B$ symmetry we can further rewrite
\begin{multline}
G^{A\bar{A}}G^{B\bar{B}}G^{C\bar{C}}  f_{\bar{A}\bar{C}}^{\bar{D}} f_{\bar{B}\bar{D}}^{\bar{E}} \eta_{AB}V_{C}^a\bar{V}_{\bar{E}}^{\bar{a}}  = 
G^{A\bar{A}}G^{B\bar{B}}G^{C\bar{C}} \left(
V_{\bar{B}}^{\bar{k}} V_{\bar{A}}^{\bar{s}}\partial_{\bar{k}}\partial_{\bar{s}}  V_{\bar{C}}^{\bar{a}} 
- 
V_{\bar{B}}^{\bar{k}} V_{\bar{C}}^{\bar{s}}\partial_{\bar{k}}  \partial_{\bar{s}}  V_{\bar{A}}^{\bar{a}}  +
V_{\bar{B}}^{\bar{k}} \partial_{\bar{k}}V_{\bar{A}}^{\bar{s}}\partial_{\bar{s}}  V_{\bar{C}}^{\bar{a}} 
-
2V_{\bar{B}}^{\bar{k}} \partial_{\bar{k}} V_{\bar{C}}^{\bar{s}}\partial_{\bar{s}}  V_{\bar{A}}^{\bar{a}} \right. \\ 
\left.   +V_{\bar{C}}^{\bar{k}} \partial_{\bar{k}} V_{\bar{B}}^{\bar{s}}\partial_{\bar{s}}  V_{\bar{A}}^{\bar{a}} 
\right)
\left(\partial_k V_A^b\partial_b V_B^k +  G^{b\bar{c}} \partial_c G_{\bar{c}k} V^c_A\partial_b V^k_B +  G^{b\bar{c}} \partial_c G_{\bar{c}k} V^c_B\partial_b V^k_A  -
	V^b_B V^k_A\partial_kG_{\bar{c}c} \partial_b G^{c\bar{c}}  \right)V_{C}^a
\end{multline}
or equivalently 
\begin{multline}
G^{A\bar{A}}G^{B\bar{B}}G^{C\bar{C}}  f_{\bar{A}\bar{C}}^{\bar{D}} f_{\bar{B}\bar{D}}^{\bar{E}} \eta_{AB} V_{C}^a\bar{V}_{\bar{E}}^{\bar{a}} = 
G^{A\bar{A}}G^{B\bar{B}} \left(
V_{\bar{B}}^{\bar{k}} V_{\bar{A}}^{\bar{s}}\partial_{\bar{k}}\partial_{\bar{s}}  G^{a\bar{a}} 
- 
V_{\bar{B}}^{\bar{k}} G^{a\bar{s}}\partial_{\bar{k}}  \partial_{\bar{s}}  V_{\bar{A}}^{\bar{a}}  +
V_{\bar{B}}^{\bar{k}} \partial_{\bar{k}}V_{\bar{A}}^{\bar{s}}\partial_{\bar{s}}  G^{a\bar{a}} -
2V_{\bar{B}}^{\bar{k}} \partial_{\bar{k}} G^{a\bar{s}}\partial_{\bar{s}}  V_{\bar{A}}^{\bar{a}}\right. \\ 
\left.  
 +G^{a\bar{k}} \partial_{\bar{k}} V_{\bar{B}}^{\bar{s}}\partial_{\bar{s}}  V_{\bar{A}}^{\bar{a}} 
\right)
\left(\partial_k V_A^b\partial_b V_B^k +  G^{b\bar{c}} \partial_c G_{\bar{c}k} V^c_A\partial_b V^k_B +  G^{b\bar{c}} \partial_c G_{\bar{c}k} V^c_B\partial_b V^k_A  -
	V^b_B V^k_A\partial_kG_{\bar{c}c} \partial_b G^{c\bar{c}}  \right).
\end{multline}
This way we obtain 
\begin{multline}
    G^{A\bar{A}}G^{B\bar{B}}G^{C\bar{C}}  f_{\bar{A}\bar{C}}^{\bar{D}} f_{\bar{B}\bar{D}}^{\bar{E}} \eta_{AB}V_{C}^a\bar{V}_{\bar{E}}^{\bar{a}}  =  G^{a\bar{s}} \nabla_k\partial_{\bar{k}} G^{b\bar{a}} \nabla_b \partial_{\bar{s}} G^{k\bar{k}} + \\ 
    \nabla_b G^{k\bar{s}} \left(
    \partial_{\bar{k}} G^{a\bar{a}} \nabla_k \partial_{\bar{s}} G^{b\bar{k}} -2 \partial_{\bar{s}} G^{a\bar{k}} \nabla_k \partial_{\bar{k}} G^{b\bar{a}} 
    +\partial_{\bar{k}}\partial_{\bar{s}} G^{a\bar{a}} \nabla_k G^{b\bar{k}}  - G^{a\bar{k}} \nabla_k \partial_{\bar{k}}\partial_{\bar{s}} G^{b\bar{a}}
    \right).
\end{multline}
Where 
\begin{equation}
    \nabla_b G^{k\bar{s}} = \partial_b G^{k\bar{s}} + G^{k\bar{c}} \partial_c G_{\bar{c}b} G^{c\bar{s}} 
\end{equation}
\begin{equation}
    \nabla_b \bar{\partial}\dots \bar{\partial}G^{k\bar{s}} = \partial_b \bar{\partial}\dots \bar{\partial}G^{k\bar{s}} + G^{k\bar{c}} \partial_c G_{\bar{c}b} \bar{\partial}\dots \bar{\partial} G^{c\bar{s}} 
\end{equation}
These computations are valid for any metric $G^{a\bar{a}}$. For K\"ahler metric they can be simplified even further. 
In particular, in this case, the covariant derivative of a metric vanishes 
\begin{equation}
        \nabla_b G^{k\bar{s}} = 0,
\end{equation}
while for the ``double'' derivative we have the following presentation
\begin{multline}
        \nabla_k \partial_{\bar{k}} G^{b\bar{a}} = \partial_k \partial_{\bar{k}} G^{b\bar{a}}  + \Gamma^{b}_{kc} \partial_{k} G^{c\bar{a}} =  
\partial_{\bar{k}}(\nabla_k  G^{b\bar{a}} ) -G^{c\bar{a}}  \partial_{\bar{k}} \Gamma^{b}_{kc}  \\  =  - R^{b}\,_{k\bar{k}c} G^{c\bar{a}}  =  R_{\bar{b}k\bar{k}c} G^{b\bar{b}}G^{c\bar{a}}  = -R_{c\bar{k}k\bar{b}} G^{b\bar{b}}G^{c\bar{a}} =-  R^{\bar{a}}\,_{\bar{k}k\bar{b}}  G^{b\bar{b}}.
\end{multline}
And in a similar way 
\begin{equation}
    \nabla_b \partial_{\bar{s}} G^{k\bar{k}}  =  -R^{k}\,_{m\bar{s}b} G^{m\bar{k}}=  -R^{k}\,_{b\bar{s}m} G^{m\bar{k}}
    =  -R_{\bar{p}b\bar{s}m} G^{m\bar{k}}G^{k\bar{p}} = -R_{\bar{s}m\bar{p}b} G^{m\bar{k}}G^{k\bar{p}}.
\end{equation}
Altogether we get 
\begin{equation}
      G^{A\bar{A}}G^{B\bar{B}}G^{C\bar{C}}  f_{\bar{A}\bar{C}}^{\bar{D}} f_{\bar{B}\bar{D}}^{\bar{E}} \eta_{AB}   = R^{\bar{a}}\,_{\bar{k}k\bar{b}} R^a\,_{m\bar{p}b} G^{m\bar{k}}G^{k\bar{p}} G^{b\bar{b}} = R^a\,_{m\bar{p}b} R^{\bar{a}m\bar{p}b},
\end{equation}
which demonstrates that even on the two-loop level 
\begin{equation}
    \boxed{\beta_3^{\rm algebra} = \beta_3^{\rm geometry}}.
\end{equation}

\section{Current-current correlation function} \label{3proof}

 In this section, we outline proof of Eqs. (19)-(21) in the main text. 

For convenience let us reformulate the problem here. 
We are interested in the computation of the following current-current correlation function
\begin{equation} \label{OPE2}
\langle \mathcal{J}_A(z)\mathcal{J}_B(0) \rangle_{S_m} \equiv - \frac{\eta_{AB}}{z^2}
\end{equation}
in the limit $m|z|\gg1$. 

The currents are given by the normal ordered expressions $\mathcal{J}_A(z) =  p_a V_A^a(\varphi) - \pi_a \partial_b V_A^a \psi^b$, while the mass deformed action reads
\begin{equation}\label{Sm}
    S_m = \underbrace{\int \frac{d^2z}{\pi}  \left(
	p_a \bar{\partial}\varphi^a -\pi_a \bar{\partial} \psi^a  + {\rm c.c.}
	\right)}_{S_0} + \underbrace{m \int    \frac{d^2z}{2\pi} G^{a\bar{a}}(\varphi)\pi_a \bar{\pi}_{\bar{a}} +m \int  \frac{d^2z}{2\pi}  G_{\bar{a}a}(\varphi) \bar{\psi}^{\bar{a}}\psi^a}_{\delta S_m}.
\end{equation}
The coefficients are chosen in such a way that at $m=0$ the OPE would read as 
 \begin{equation}
 	p_a(z)\varphi^b(w)\Big|_{m=0} = \frac{\delta_a^b}{z-w} + {\rm reg}, \qquad \pi_a(z)\psi^b(w)\Big|_{m=0} = \frac{\delta_a^b}{z-w} + {\rm reg}. 
 \end{equation}
We are going to prove that 
\begin{equation}\label{etaAB}
    \eta_{AB} =\nabla_a V_A^b\nabla_b V_B^a= 
 \partial_a V_A^b\partial_b V_B^a +  G^{b\bar{a}} \partial_c G_{\bar{a}a} V^c_A\partial_b V^a_B+  G^{b\bar{a}} \partial_c G_{\bar{a}a} V^c_B\partial_b V^a_A  -
	V^b_B V^a_A\partial_aG_{\bar{c}c} \partial_b G^{c\bar{c}}.
\end{equation}
First let us notice that in the correlators that do not involve $p$, one can easily ignore $\varphi$ dependence in $G_{a\bar{a}}$ and $G^{a\bar{a}}$
in $S_m$. 
In particular, the fermions correlators could be computed explicitly (we drop $S_m$ subscript in the correlators), 
\begin{equation}
	\langle \pi_a(z) \psi^b(0)	
	\rangle  = \delta_a^b \frac{mz}{|z|} K_1(m|z| ),\qquad 
		\langle \bar{\pi}_{\bar{a}}(z) \bar{\psi}^{\bar{b}}(0)	
	\rangle =\delta_{\bar{a}}^{\bar{b}}\frac{m\bar{z}}{|z|} K_1(m|z| )
\end{equation}
 \begin{equation}
 	\langle \pi_a(z) \bar{\pi}_{\bar{a}}(0)	
 \rangle  = m G_{\bar{a}a}K_0(m|z|),\qquad 
 \langle \bar{\psi}^{\bar{a}}(z)\psi^a(0)	
 \rangle  = m G^{a\bar{a}}K_0(m|z|).
 \end{equation}
 Here $K_0$ and $K_1$ are the modified Bessel function of the second kind. 
 The rest of fermionic two-point functions are zero. 
 All these correlators are exponentially small for $m|z|\gg1$. 
 This way, the fermion-fermion current terms do not contribute in the correlator \eqref{OPE2}  
 \begin{equation}
 	\langle \pi_a \partial_b V_A^a \psi^b(z)  \pi_c \partial_n V_B^c \psi^n(0) \rangle  = O(e^{-m|z|}).
 \end{equation}
 Now let us have a look on mixed terms. 
In this case, the field $p$ from the bosonic part of the current can be contracted with the $\varphi$ in the deformation terms in the mass part of $S_m$ \eqref{Sm}. 
Namely, 
\begin{equation}
	\langle p_aV_A^a(z)  \pi_c \partial_n V_B^c \psi^n(0) \delta S_m\rangle = 2 V_A^a\partial_nV_B^cG^{n\bar{a}} \partial_a G_{\bar{a}c} \int \frac{d^2w}{2\pi} \frac{m^3}{z-w} \frac{\bar{w}}{|w|} K_1(m|w|) K_0(m|w|) 
\end{equation}
 Notice that the above integral does not have UV divergences, to compute it we use the  radial coordinates and 
rescale the absolute value $|w|$
 \begin{equation}
 	  \int\limits_0^\infty d|w| \int\limits_{-\pi}^\pi \frac{d\varphi}{\pi} \frac{m |w|e^{-i\varphi }}{z-e^{i\varphi}|w|/m} K_0(|w|)K_1(|w|)  =  \frac{2}{z^2}   \int\limits_0^{m|z|} |w|^2 K_0(|w|)K_1(|w|)   d|w| 
 \end{equation}
 For $ m |z| \gg 1$ we can replace the upper limit to infinity, which will yield us some constant 
 \begin{equation}
  \int\limits_0^{\infty} |w|^2K_0(|w|)K_1(|w|)    d|w|  = \frac{1}{2}.
 \end{equation}
 This gives 
 \begin{equation}
 		\langle p_aV_A^a(z)  \pi_c \partial_n V_B^c \psi^n(0) \delta S_m\rangle_{S_0}  \overset{m|z|\gg 1}{=} \frac{V_A^a\partial_nV_B^c G^{n\bar{a}} \partial_a G_{\bar{a}c}}{z^2}
 \end{equation}
This way, we recover $V V'$ terms in \eqref{etaAB}.
 
 Now let us turn to the boson-boson contributions. 
To simplify computations we consider a small deformation on top of the constant metric $G_{\bar{a}a} = [G_0]_{\bar{a}a}+ \delta G_{\bar{a}a} $, 
and correspondingly ${G^{a\bar{a}} = [G_0]^{a\bar{a}}- \delta G^{\bar{a}a}}$, where 
\begin{equation}
\delta G^{\bar{a}a} \equiv G_0^{\bar{a}c}\delta G_{\bar{c}c} G_0^{\bar{c}a} + O(\delta G^2).
\end{equation} 
Quadratic in $\delta G_{\bar{a}a}$ terms are 
 \begin{equation}
	\left\langle 
p_a V_A^a(z) p_b V^b_B(0) \frac{(\delta S_m)^2}{2}
\right\rangle = I_{\pi\pi} + I_{\pi\psi} + I_{\psi\psi}
 \end{equation}
 where 
 \begin{equation}
 	I_{\pi\pi} = \frac{m^2}{2} \int \frac{d^2w_1}{2\pi}	\int \frac{d^2w_2}{2\pi}  	\left\langle 
 	p_a V_A^a(z) p_b V^b_B(0) \delta G^{c\bar{c}}\pi_c \bar{\pi}_{\bar{c}}(w_1) \delta G^{k\bar{k}}\pi_k \bar{\pi}_{\bar{k}}(w_2)  
 	\right\rangle_{G_0}
 \end{equation}
  \begin{equation}
 	I_{\psi\psi} = \frac{m^2}{2} \int \frac{d^2w_1}{2\pi}	\int \frac{d^2w_2}{2\pi}  	\left\langle 
 	p_a V_A^a(z) p_b V^b_B(0) \delta G_{\bar{c}c}\bar{\psi}^{\bar{c}}\psi^c (w_1) \delta G_{\bar{k}k}\bar{\psi}^{\bar{k}}\psi^k(w_2)  
 	\right\rangle_{G_0}
 \end{equation}
   \begin{equation}
 	I_{\pi\psi} = -m^2\int \frac{d^2w_1}{2\pi}	\int \frac{d^2w_2}{2\pi}  	\left\langle 
 	p_a V_A^a(z) p_b V^b_B(0) \delta G^{c\bar{c}}\pi_c \bar{\pi}_{\bar{c}}(w_1) \delta G_{\bar{k}k}\bar{\psi}^{\bar{k}}\psi^k(w_2)  
 	\right\rangle_{G_0}
 \end{equation}
 Here the correlators are computed with the action $S_m$ with metric $G_0$. 
 Now let us focus on terms that are proportional to $V^a_A V^b_B $ i.e. where vectors fields are not differentiated, meaning that the  fields $p$ are contracted with $\varphi$ in the perturbation. 
 Let us additionally introduce 
 \begin{equation}
 	h_a^c = G_0^{\bar{c}c}\delta G_{\bar{c}a}
 \end{equation}
 then we obtain 
 \begin{multline}
 		\left\langle 
 	p_a V_A^a(z) p_b V^b_B(0) \frac{(\delta S)^2}{2}
 	\right\rangle\Big|_{VV} = - V_A^aV_B^b \int \frac{d^2w_1}{2\pi}	  \int \frac{d^2w_2}{2\pi} m^4 \left[ K_1(m|w_1-w_2|)^2 + K_0(m|w_1-w_2|)^2\right] \times  \\
 	\left(
 	\frac{\partial_ah_{c}^d\partial_bh_d^c}{(z-w_1)w_2} +
 	\frac{h_d^c \partial_a\partial_b h_{c}^d}{(z-w_1)w_1} + ( w_1, \,a \leftrightarrow w_2, \,b )
 	\right) 	 
 \end{multline}
 Identically we can present this expression as 
  \begin{equation}\label{ff}
 	\left\langle 
 	p_a V_A^a(z) p_b V^b_B(0) \frac{(\delta S)^2}{2}
 	\right\rangle\Big|_{VV}  = V_A^aV_B^b \partial_ah_{c}^d\partial_bh_d^c C_1 \\ 
 	- V_A^aV_B^b \partial_a\partial_b (h_{c}^dh_d^c)C_2 
 \end{equation}
 where 
 \begin{multline}
 	C_1 = \int \frac{d^2w_1}{2\pi}	\int \frac{d^2w_2}{2\pi}  \frac{2m^4w_{21}(K_1(m|w_{21}|)^2+K_0(m|w_{21}|)^2)}{(z-w_1)w_1w_2} = \\
 	\int \frac{d^2w_1}{2\pi}	\int \frac{d^2w_2}{2\pi}  \frac{2m^4w_{2}(K_1(m|w_{2}|)^2+K_0(m|w_{2}|)^2)}{(z-w_1)w_1(w_2-w_1)}
 \end{multline}
 \begin{equation}
 	C_2 = \int \frac{d^2w_1}{2\pi}	\int \frac{d^2w_2}{2\pi}  \frac{2m^4(K_1(m|w_{21}|)^2+K_0(m|w_{21}|)^2)}{(z-w_1)w_1}  = \int \frac{d^2w_1}{2\pi}	\frac{C_3}{(z-w_1)w_1}  
 \end{equation}
 with $w_{12} = w_1 -w_2$, and $C_3$ equals to
  \begin{equation}
 	C_3 = 	\int \frac{d^2w_2}{2\pi}2m^4(K_1(m|w_{2}|)^2+K_0(m|w_{2}|)^2)
 \end{equation}
 This integral is divergent logarithmically, but luckily it can be regularized by adding counter terms to the action 
 \begin{equation}
 	 S_m \to S_m - C_3\int  h_{c}^dh_d^c(z) d^2z
 \end{equation}
 Notice this regularization is also needed for $VV'$ terms. 
 The constant $C_1$ is finite and can be computed as follows. First, we perform integration over the angle of $w_2$
 \begin{equation}
	\int \frac{d^2w_2}{2\pi} \frac{w_2f(|w_2|)}{w_2-w_1} =  \int\limits_{|w_1|}^\infty |w_2|f(|w_2|)d|w_2|
\end{equation}
and then similarly over angle of $w_1$ 
 \begin{equation}
	\int \frac{d^2w_1}{2\pi}   \frac{f(|w_1|)}{(z-w_1)w_1} = \frac{1}{z^2}\int\limits_{0}^{|z|} |w_1| f(|w_1|) d|w_1|,
\end{equation}
which leaves us with the two-dimensional integral over absolute values (we also rescale $|w_i|\to |w_i|/m$)
\begin{equation}
C_1 = \frac{2}{z^2}\int\limits_{0}^{m|z|}  \, d|w_1|\, |w_1|   \int\limits_{|w_1|}^\infty |w_2| \left(K_1(|w_2|)^2+K_0(|w_2|)^2 \right)d|w_2|
\end{equation}
For $m|z| \gg1$ we replace the upper limit with $\infty$ and then exchange integration over $|w_1|$ and $|w_2|$, in the end we arrive at 
\begin{equation}
	C_1 =  \frac{1}{z^2} \int\limits_{0}^\infty |w_2|^3 \left(K_1(|w_2|)^2+K_0(|w_2|)^2 \right)d|w_2| = \frac{1}{z^2}.
\end{equation}
This way, without $C_2$ and with $C_1=1/z^2$ from \eqref{ff} we recover quadratic terms in the expansion of \eqref{etaAB} around the constant metric $G_0$ on terms without derivative of the vector fields.
